\documentstyle[floats,prl,aps]{revtex}
\begin{document}
\renewcommand{\thefootnote}{\fnsymbol{footnote}}
\draft
\title{\large\bf 
  New integrable boundary conditions for the $q$-deformed supersymmetric 
  $U$ model and Bethe ansatz equations}

\author{Yao-Zhong Zhang \footnote {E-mail: yzz@maths.uq.edu.au}
             and 
        Huan-Qiang Zhou \footnote {E-mail: hqz@maths.uq.edu.au}}

\address{      Department of Mathematics,University of Queensland,
		     Brisbane, Qld 4072, Australia}

\maketitle

\vspace{10pt}

\begin{abstract}
New classes of integrable boundary conditions for the $q$-deformed 
(or two-parameter) supersymmetric $U$ model are presented.
The boundary systems are solved by using
the coordinate space Bethe ansatz technique and Bethe ansatz equations
are derived. 
\end{abstract}


%************************** Text Begins here ******************************

%  Greek letters

\def\a{\alpha}
\def\b{\beta}
\def\d{\delta}
\def\e{\epsilon}
\def\g{\gamma}
\def\k{\kappa}
\def\l{\lambda}
\def\o{\omega}
\def\t{\theta}
\def\s{\sigma}
\def\D{\Delta}
\def\L{\Lambda}

% Shorthands for \begin{equation} and the like

\def\beq{\begin{equation}}
\def\eeq{\end{equation}}
\def\bea{\begin{eqnarray}}
\def\eea{\end{eqnarray}}
\def\ba{\begin{array}}
\def\ea{\end{array}}
\def\no{\nonumber}
\def\le{\langle}
\def\re{\rangle}
\def\lt{\left}
\def\rt{\right}

\newcommand{\reff}[1]{eq.~(\ref{#1})}

%\newpage
\vskip.3in

Integrable correlated electron systems (see, e.g. \cite{Ess94})
with open boundary conditions
constitute an important class of physically significant models
which are exactly solvable by means of the Bethe ansatz
technique and the boundary quantum inverse scattering method 
(QISM) \cite{Skl88,Mez91}. 
In this letter, we determine integrable boundary conditions for
the $q$-deformed (or two-parameter) supersymmetric
$U$ model \cite{Bra95,Bar95,Gou96} of strongly correlated electrons. 
We present  four
classes of boundary conditions for this model, all of which are
integrable by the supersymmetric boundary QISM formulated recently in
\cite{Bra97}. We solve the boundary systems using the coordinate
Bethe ansatz method and derive the Bethe ansatz equations for all 
four cases.

Let $c_{j,\s}$ and $c_{j,\s}^{\dagger}$ denote fermionic creation and
annihilation operators with spin $\s$ at
site $j$, which satisfy the anti-commutation relations given by
$\{c_{i,\s}^\dagger, c_{j,\tau}\}=\d_{ij}\d_{\s\tau}$, where 
$i,j=1,2,\cdots,L$ and $\s,\tau=\uparrow,\;\downarrow$. We consider 
open-boundary $q$-deformed supersymmetric $U$ models with the 
Hamiltonian of the form:
\beq
H=\sum _{j=1}^{L-1} H_{j,j+1}^Q + B_{lt} +B_{rt},\label{h}\label{hamiltonian}
\eeq
where $B_{lt}~ (B_{rt})$ stands for left (right) boundary term whose explicit  
forms are spelled out below, and
$H_{j,j+1}^Q$ is the bulk Hamiltonian density of the $q$-deformed
supersymmetric $U$ model \cite{Bar95,Gou96}
\bea
H_{j,j+1}^Q&=&-\sum _{\sigma}(c^{\dagger}_{j\sigma}c_{j+1\sigma}+h.c.)
  \exp(-\frac {1}{2}(\eta -{\rm sign}(\s) \g)n_{j,-\sigma}-\frac {1}{2}
  (\eta + {\rm sign}(\s) \g)n_{j+1,-\sigma})\no\\
& &+\frac {U}{2}(n_{j\uparrow}n_{j\downarrow}
  +n_{j+1\uparrow}n_{j+1\downarrow})\no\\
& &+t_p(c^{\dagger}_{j\uparrow}c^{\dagger}_{j\downarrow}c_{j+1\downarrow}
  c_{j+1\uparrow}+h.c.)+e^{-2a}\; n_{j}+e^{2a}\;n_{j+1}.\label{h-density}
\eea
Here ${\rm sign}(\s)=1$ for $\s=\uparrow$ and $-1$ for $\s=\downarrow$,
$n_{j\s}$ is the density operator
$n_{j\s}=c_{j\s}^{\dagger}c_{j\s}$,
$n_j=n_{j\uparrow}+n_{j\downarrow}$, and $\eta,~\g$ are two free parameters; 
and
\bea
t_p&=&\frac {U}{2} =
  [2e^{-\eta}(\cosh \eta -\cosh \g)]^{\frac {1}{2}},\no\\
a&=& \frac {1}{4} \{ \ln [\frac {\sinh \frac {1}{2}(\eta +\g)}
 {\sinh \frac {1}{2}(\eta -\g)}]-\g \}.
%\mu& =&\sqrt{ e^{\g}\frac {\sinh (\eta -\g)/2}{\sinh (\eta +\g)/2}}.
\eea

Let us point out that in the limit of $\eta=\g$ (first discarding
the chemical potential terms and then taking the
limit), the bulk model Hamiltonian (\ref{h-density}) reduces to that
of Bariev \cite{Bar91}. This implies that, up to the chemical
potential terms,  the bulk model (\ref{h-density}) is a 
generalization of the Bariev's model \cite{Bar91}. 

We propose the following four classes of boundary conditions:
\bea
{\rm Case~ A:}~~ &&B_{lt}=\frac {\sinh 2a}
  {\sinh \frac {\g(2-\xi_-)}{2}}
\lt(\frac {\sinh \g}{\sinh \frac {\g \xi_-}{2}}
  n_{1\uparrow}n_{1\downarrow}
  -e^{-\g (1-\frac {\xi_-}{2})}n_1\rt),\no\\
&&B_{rt}=\frac {\sinh 2a}
  {\sinh \frac {\g(2-\xi_+)}{2}}
\lt(\frac {\sinh \g}{\sinh \frac {\g \xi_+}{2}}
  n_{L\uparrow}n_{L\downarrow}
  -e^{\g (1-\frac {\xi_+}{2})}n_L\rt);\label{boundary11}\\
{\rm Case~B:}~~ &&B_{lt}=\frac{\sinh 2a}{\sinh(\g\zeta_-+a)}
  e^{\g\zeta_-+a}\;n_{1\downarrow},~~~~~
B_{rt}=-\frac{\sinh 2a}{\sinh(\g\zeta_+-a)}
		e^{\g\zeta_+-a}\;n_{L\downarrow};\label{boundary22}\\
{\rm Case~ C:}~~ &&B_{lt}=\frac {\sinh 2a}
  {\sinh \frac {\g(2-\xi_-)}{2}}
\lt(\frac {\sinh \g}{\sinh \frac {\g \xi_-}{2}}
  n_{1\uparrow}n_{1\downarrow}
  -e^{-\g (1-\frac {\xi_-}{2})}n_1\rt),\no\\
&&B_{rt}=-\frac{\sinh 2a}{\sinh(\g\zeta_+-a)}
		e^{\g\zeta_+-a}\;n_{L\downarrow};\label{boundary12}\\
{\rm Case~D:}~~ &&B_{lt}=\frac{\sinh 2a}{\sinh(\g\zeta_-+a)}
  e^{\g\zeta_-+a}\;n_{1\downarrow},\no\\
&&B_{rt}=\frac {\sinh 2a}
  {\sinh \frac {\g(2-\xi_+)}{2}}
\lt(\frac {\sinh \g}{\sinh \frac {\g \xi_+}{2}}
  n_{L\uparrow}n_{L\downarrow}
  -e^{\g (1-\frac {\xi_+}{2})}n_L\rt),\label{boundary21}
\eea
where $\xi_{\pm}$ and $\zeta_\pm$ are some parameters describing the
boundary effects. 

Quantum integrability of the system defined
by Hamiltonian (\ref{hamiltonian}) with all four boundary conditions
can be established as follows by means of the supersymmetric boundary QISM 
\cite{Bra97} in which the above Case A is treated in details as an
example. We first search for boundary K-matrices which satisfy 
the graded reflection equations \cite{Bra97}:
\bea
&&R_{12}(u_1-u_2)\stackrel {1} {K}_-(u_1) R_{21}(u_1+u_2)
\stackrel {2}{K}_-(u_2)= 
\stackrel {2}{K}_-(u_2) R_{12}(u_1+u_2)
\stackrel {1}{K}_-(u_1) R_{21}(u_1-u_2),\no\\ 
&&R_{12}(-u_1+u_2)\stackrel {1}{K_+}(u_1)\stackrel{1}{M^{-1}} 
R_{21}(-u_1-u_2+2)\stackrel{1}{M}
\stackrel {2}{K_+}(u_2)\no\\
&&~~~~~~~~~~~~~~~~~~~~~~~~~~ =\stackrel{1}{M} 
\stackrel {2}{K_+}(u_2) R_{12}(-u_1-u_2+2)\stackrel{1}{M^{-1}}
\stackrel {1}{K_+}(u_1) R_{21}(-u_1+u_2),\label{RE-with-cu}
\eea
where $R(u)$ is the R-matrix of the
$q$-deformed supersymmetric $U$ model \cite{Gou96}, $R_{21}(u)=
P_{12}R_{12}(u)P_{12}$ with $P$ being the graded permutation operator and
$M={\rm diag} (1,1,e^{2\g},e^{2\g})$ is the so-called crossing matrix.
There are two different diagonal boundary K-matrices, $K^I_-(u),~
K^{II}_-(u)$, which solve the first reflection equation in
(\ref{RE-with-cu}) \cite{Arn97}:
\bea
K^I_-(u)&=&  \frac {1}{ \sinh \frac {\g \xi_-}{2}\sinh \frac {\g(\xi_--2)}{2}}
  {\rm diag} \lt(A_-(u),B_-(u),B_-(u),C_-(u)\rt),\label{k-1}\\
K^{II}_-(u)&=&{\rm diag}\lt(\bar{A}_-(u),\bar{A}_-(u),\bar{B}_-(u),
  \bar{B}_-(u)\rt),\label{k-2} 
\eea
where
\bea
A_-(u)&=&e^{\g u} \sinh  \frac {\g (\xi_-+u)}{2}\sinh 
   \frac {\g (u-2+\xi_-)}{2},\no\\
B_-(u)&=& \sinh \frac {\g (\xi_--u)}{2}\sinh \frac {\g (u-2+\xi_-)}{2},\no\\
C_-(u)&=&e^{-\g u} \sinh \frac {\g (\xi_--u)}{2}\sinh 
   \frac {\g (-u-2+\xi_-)}{2},\no\\
\bar{A}_-(u)&=&1+\frac{e^{-\g u}-1}{2\sinh(\g\zeta_-+a)}
   e^{\g\zeta_-+a},\no\\
\bar{B}_-(u)&=&1+\frac{e^{\g u}-1}{2\sinh(\g\zeta_-+a)}
   e^{\g\zeta_-+a}.
\eea
The corresponding K-matrices $K^{I\,(II)}_+(u)$ which obey 
the second reflection
equation in (\ref{RE-with-cu}) are derived by isomorphism:
\beq
K^I_+(u)=M K^I_-(-u+1),~~~~~~K^{II}_+(u)=MK^{II}_-(-u+1)
\eeq
 
We form the boundary transfer matrix $t(u)$:
\beq
t(u)=str[K_+(u)T_-(u)K_-(u)T^{-1}_-(-u)],~~~~~~
  T_-(u) = R_{0L}(u) \cdots R_{01}(u).
\eeq
Since $K_\pm(u)$  can be taken as $K^I_\pm(u)$ or $K^{II}_\pm(u)$,
respectively, we have four possible choices of boundary transfer matrices:
\bea
t^A(u)&=&str[K^I_+(u)T_-(u)K^I_-(u)T^{-1}_-(-u)],\no\\
t^B(u)&=&str[K^{II}_+(u)T_-(u)K^{II}_-(u)T^{-1}_-(-u)],\no\\
t^C(u)&=&str[K^{II}_+(u)T_-(u)K^I_-(u)T^{-1}_-(-u)],\no\\
t^D(u)&=&str[K^I_+(u)T_-(u)K^{II}_-(u)T^{-1}_-(-u)],\label{t-matrices}
\eea
which reflects the fact that 
the boundary conditions on the left end and on the right end of
the open lattice chain are independent.
Then it can be shown \cite{Bra97} that Hamiltonians corresponding to 
all four boundary 
conditions can be embedded into the above four boundary transfer matrices,
respectively. We thus arrive at the four possible
cases (\ref{boundary11}--{\ref{boundary21}), all of which are 
compatible with the bulk integrability. 

We now solve the above boundary systems by using
the coordinate space Bethe ansatz method. Following \cite{Asa96,Bra97},
we assume that the eigenfunction of Hamiltonian (\ref{hamiltonian}) 
takes the form
\bea
| \Psi \rangle& =&\sum _{\{(x_j,\s_j)\}}\Psi _{\s_1,\cdots,\s_N}
  (x_1,\cdots,x_N)c^\dagger
  _{x_1\s_1}\cdots c^\dagger_{x_N\s_N} | 0 \rangle,\no\\
\Psi_{\s_1,\cdots,\s_ N}(x_1,\cdots,x_N)
&=&\sum _P \e _P A_{\s_{Q1},\cdots,\s_{QN}}(k_{PQ1},\cdots,k_{PQN})
\exp (i\sum ^N_{j=1} k_{P_j}x_j),
\eea
where the summation is taken over all permutations and negations of
$k_1,\cdots,k_N,$ and $Q$ is the permutation of the $N$ particles such that
$1\leq   x_{Q1}\leq   \cdots  \leq  x_{QN}\leq   L$.
The symbol $\e_P$ is a sign factor $\pm1$ and changes its sign
under each 'mutation'. Substituting the wavefunction into  the
eigenvalue equation $ H| \Psi  \rangle = E | \Psi \rangle $,
one gets
\bea
A_{\cdots,\s_j,\s_i,\cdots}(\cdots,k_j,k_i,\cdots)&=&S_{ij}(
    k_i,k_j)
    A_{\cdots,\s_i,\s_j,\cdots}(\cdots,k_i,k_j,\cdots),\no\\
A_{\s_i,\cdots}(-k_j,\cdots)&=&s^L(k_j;p_{1\s_i})A_{\s_i,\cdots}
    (k_j,\cdots),\no\\
A_{\cdots,\s_i}(\cdots,-k_j)&=&s^R(k_j;p_{L\s_i})A_{\cdots,\s_i}(\cdots,k_j),
\eea
where $S_{ij}(k_i,k_j) $ are
the two-particle scattering matrices,
\bea
S_{ij}(k_i,k_j)^{11}_{11}&=&S_{ij}(k_i,k_j)^{22}_{22}=1,\no\\
S_{ij}(k_i,k_j)^{12}_{12}&=&S_{ij}(k_i,k_j)^{21}_{21}=
  \frac{\sin(\l_i-\l_j)}{\sin (\l_i-\l_j-i \g)},\no\\
S_{ij}(k_i,k_j)^{12}_{21}&=&
  e^{-i\;(\l_i-\l_j)}\frac{\sin i\g}{\sin(\l_i-\l_j-i\g)},\no\\
S_{ij}(k_i,k_j)^{21}_{12}&=&
  e^{i\;(\l_i-\l_j)}\frac{\sin i\g}{\sin(\l_i-\l_j-i\g)}\label{s-matrix}
\eea
with $\l_j$ being suitable particle
rapidities related to the quasi-momenta $k_j$ of the electrons by
\cite{Bar95}
\beq
k(\l)=2 \arctan (\coth a \tan \l)
\eeq
and $s^L(k_j;p_{1\s_i})$,
 $~s^R(k_j;p_{L\s_i})$ are the boundary scatering matrices,
\bea
s^L(k_j;p_{1\s_i})&=&\frac {1-p_{1\s_i}e^{ik_j}}
{1-p_{1\s_i}e^{-ik_j}},\no\\
s^R(k_j;p_{L\s_i})&=&\frac {1-p_{L\s_i}e^{-ik_j}}
{1-p_{L\s_i}e^{ik_j}}e^{2ik_j(L+1)}
\eea
with $p_{1\s_i}$ and $p_{L\s_i}$ being given by the following formulae,
corresponding to the four cases, respectively,
\bea
{\rm Case~A:}~~
&&p_{1\uparrow}=p_{1\downarrow}\equiv p_1=-e^{2a}- 
  \frac {\sinh 2a}
  {\sinh \frac {\g(2-\xi_-)}{2}}
   e^{-\g (1-\frac {\xi_-}{2})},\no\\
&&p_{L\uparrow}=p_{L\downarrow}\equiv p_L=-e^{-2a}-
  \frac {\sinh 2a}
  {\sinh \frac {\g(2-\xi_+)}{2}}
   e^{\g (1-\frac {\xi_+}{2})};\label{pa}\\
{\rm Case~B:}~~&&p_{1\uparrow}=-e^{2a},~~~~~p_{1\downarrow}=-e^{2a}
   +\frac{\sinh 2a}{\sinh(\g\zeta_-+a)}e^{\g\zeta_-+a},\no\\
&&p_{L\uparrow}=-e^{-2a},~~~~~p_{L\downarrow}=-e^{-2a}
   -\frac{\sinh 2a}{\sinh(\g\zeta_+-a)}e^{\g\zeta_+-a};\label{pb}\\
{\rm Case~C:}~~
&&p_{1\uparrow}=p_{1\downarrow}\equiv p_1=-e^{2a}- 
  \frac {\sinh 2a}
  {\sinh \frac {\g(2-\xi_-)}{2}}
   e^{-\g (1-\frac {\xi_-}{2})},\no\\
&&p_{L\uparrow}=-e^{-2a},~~~~~p_{L\downarrow}=-e^{-2a}
   -\frac{\sinh 2a}{\sinh(\g\zeta_+-a)}e^{\g\zeta_+-a};\label{pc}\\
{\rm Case~D:}~~&&p_{1\uparrow}=-e^{2a},~~~~~p_{1\downarrow}=-e^{2a}
   +\frac{\sinh 2a}{\sinh(\g\zeta_-+a)}e^{\g\zeta_-+a},\no\\
&&p_{L\uparrow}=p_{L\downarrow}\equiv p_L=-e^{-2a}-
  \frac {\sinh 2a}
  {\sinh \frac {\g(2-\xi_+)}{2}}
   e^{\g (1-\frac {\xi_+}{2})}.\label{pd}
\eea

As is seen above, the two-particle S-matrix (\ref{s-matrix}) is 
nothing but the R-matrix of the six-vertex model in the homogenous
gradation and thus
satisfies the quantum Yang-Baxter equation (QYBE),
\beq
S_{ij}(k_i,k_j)S_{il}(k_i,k_l)S_{jl}(k_j,k_l)=
S_{jl}(k_j,k_l)S_{il}(k_i,k_l)S_{ij}(k_i,k_j).
\eeq
It can be checked that the boundary scattering matrices $s^L$ and $s^R$ 
obey the reflection equations:
\bea
&&S_{ji}(-k_j,-k_i)s^L(k_j;p_{1\s_j})S_{ij}(-k_i,k_j)s^L(k_i;p_{1\s
  _i})\no\\
&&~~~~~~~~~~~~~~~~~~=s^L(k_i;p_{1\s _i})S_{ji}(-k_j,k_i)s^L(k_j;p_{1\s _i})
  S_{ij}(k_i,k_j),\no\\
&&S_{ji}(-k_j,-k_i)s^R(k_j;p_{L\s_j})S_{ij}(k_i,-k_j)s^R(k_i;p_{L\s
  _i})\no\\
&&~~~~~~~~~~~~~~~~~~= s^R(k_i;p_{L\s _i})S_{ji}(k_j,-k_i)s^R(k_j;p_{L\s _i})
  ;p_{\s_i})S_{ji}(k_j,k_i).\label{reflection-e}
\eea
This is seen as follows. One introduces the notation
\beq
s(k;p)=\frac  {1-pe^{-ik}}{1-p e^{ik}}.
\eeq
Then the boundary scattering matrices $s^L(k_j;p_{1\s_i})$,
 $~s^R(k_j;p_{L\s_i})$ can be written as, corresponding to the four
cases, respectively,
\bea
{\rm Case~A:}~~&&s^L(k_j;p_{1\s_i})=s(-k_j;p_1)I,\no\\
&&s^R(k_j;p_{L\s_i})=e^{ik_j2(L+1)}s(k_j;p_L)I;\label{sa}\\
{\rm Case~B:}~~&&s^L(k_j;p_{1\s_i})=s(-k_j;p_{1\downarrow})\lt(
\begin{array}{cc}
e^{2i\l_j}\frac{\sin(\frac{\g\zeta_-}{i}+\l_j)}
	       {\sin(\frac{\g\zeta_-}{i}-\l_j)} & 0\\
0& 1
\end{array}
\rt),\no\\
&&s^R(k_j;p_{L\s_i})=e^{ik_j2(L+1)}s(k_j;p_{L\downarrow})\lt(
\begin{array}{cc}
e^{-2i\l_j}\frac{\sin(\frac{\g\zeta_+}{i}-\l_j)}
	       {\sin(\frac{\g\zeta_+}{i}+\l_j)} & 0\\
0& 1
\end{array}
\rt);\label{sb}\\
{\rm Case~C:}~~&&s^L(k_j;p_{1\s_i})=s(-k_j;p_1)I,\no\\
&&s^R(k_j;p_{L\s_i})=e^{ik_j2(L+1)}s(k_j;p_{L\downarrow})\lt(
\begin{array}{cc}
e^{-2i\l_j}\frac{\sin(\frac{\g\zeta_+}{i}-\l_j)}
	       {\sin(\frac{\g\zeta_+}{i}+\l_j)} & 0\\
0& 1
\end{array}
\rt);\label{sc}\\
{\rm Case~D:}~~&&s^L(k_j;p_{1\s_i})=s(-k_j;p_{1\downarrow})\lt(
\begin{array}{cc}
e^{2i\l_j}\frac{\sin(\frac{\g\zeta_-}{i}+\l_j)}
	       {\sin(\frac{\g\zeta_-}{i}-\l_j)} & 0\\
0& 1
\end{array}
\rt),\no\\
&&s^R(k_j;p_{L\s_i})=e^{ik_j2(L+1)}s(k_j;p_L)I.\label{sd}
\eea
Here $I$ stands for $2\times 2$ identity matrix and 
$p_{1\downarrow},~p_{L\downarrow}$ are the ones given in (\ref{pb}).
We immediately see that (\ref{sa}) are the trivial solutions of the reflection
equations (\ref{reflection-e}), whereas (\ref{sb}) are the diagonal
solutions \cite{Skl88,deV93}. 

The diagonalization of Hamiltonian (\ref{hamiltonian}) reduces 
to solving  the following matrix  eigenvalue equation
\beq
T_jt= t,~~~~~~~j=1,\cdots,N,
\eeq
where $t$ denotes an eigenvector on the space of the spin variables
and $T_j$ takes the form
\beq
T_j=S_j^-(k_j)s^L(-k_j;p_{1\s_j})R^-_j(k_j)R^+_j(k_j)
    s^R(k_j;p_{L\s_j})S^+_j(k_j)
\eeq
with
\bea
S_j^+(k_j)&=&S_{j,N}(k_j,k_N) \cdots S_{j,j+1}(k_j,k_{j+1}),\no\\
S^-_j(k_j)&=&S_{j,j-1}(k_j,k_{j-1})\cdots S_{j,1}(k_j,k_1),\no\\
R^-_j(k_j)&=&S_{1,j}(k_1,-k_j)\cdots S_{j-1,j}(k_{j-1},-k_j),\no\\
R^+_j(k_j)&=&S_{j+1,j}(k_{j+1},-k_j)\cdots S_{N,j}(k_N,-k_j).
\eea
This problem can  be solved using the algebraic Bethe ansatz method.
The Bethe ansatz equations for all the four cases are 
\bea
e^{ik_j2(L+1)}F(k_j;p_{1\s},p_{L\s})
&=&\prod ^{M}_{\a =1}\frac {\sin 
  (\l_j-\Lambda _{\a}-\frac {i\g}{2})}{\sin
  (\l_j-\Lambda _{\a}+\frac {i\g}{2})}\frac{\sin
  (\l_j+\Lambda _{\a}-\frac {i\g}{2})}
  {\sin (\l_j+\Lambda _{\a}+\frac {i\g}{2})},\no\\
\prod ^{N}_{j=1}
  \frac {\sin (\Lambda _{\a}-\l _j+\frac{i\g}{2})}
  {\sin (\Lambda _{\a}-\l _j-\frac{i\g}{2})}
  \frac {\sin (\Lambda _{\a}+\l _j+\frac{i\g}{2})}
   {\sin (\Lambda _{\a}+\l _j-\frac{i\g}{2})}
&=& G(\L_\a;p_{1\s},p_{L\s})
\prod ^{M}_{\stackrel {\b =1}{\b \neq \a}}
  \frac {\sin (\Lambda _{\a}-\Lambda _{\b}
   +i\g)}{\sin (\Lambda _{\a}-\Lambda _{\b}
  -i\g)}\frac {\sin (\Lambda _{\a}+\Lambda _{\b}
  +i\g)}{\sin (\Lambda _{\a}+\Lambda _{\b}-i\g)},
\eea
where
\bea
F(k_j;p_{1\s},p_{L\s})&=&\lt\{
\begin{array}{ll}
s(k_j;p_1)s(k_j;p_L) & ({\rm Case~A})\\
s(k_j;p_{1\uparrow})s(k_j;p_{L\uparrow})    & ({\rm Case~B})\\
s(k_j;p_1)s(k_j;p_{L\uparrow})    & ({\rm Case~C})\\
s(k_j;p_{1\uparrow})s(k_j;p_L)    & ({\rm Case~D})\;,
\end{array}
\rt.\no\\
G(\L_\a;p_{1\s},p_{L\s})&=&\lt\{
\begin{array}{ll}
1 & ({\rm Case~A})\\
\frac{\sin(\L_\a-i\g\zeta_++i\frac{\g}{2})\sin(\L_\a-i\g\zeta_-+
   i\frac{\g}{2})}
  {\sin(\L_\a+i\g\zeta_+-i\frac{\g}{2})\sin(\L_\a+i\g\zeta_--
   i\frac{\g}{2})} & ({\rm Case~B})\\
-\frac{\sin(\L_\a-i\g\zeta_++i\frac{\g}{2})}
 {\sin(\L_\a+i\g\zeta_+-i\frac{\g}{2})} & ({\rm Case~C})\\
-\frac{\sin(\L_\a-i\g\zeta_-+i\frac{\g}{2})}
 {\sin(\L_\a+i\g\zeta_--i\frac{\g}{2})} & ({\rm Case~D})\;,
\end{array}
\rt.
\eea
where again $p_{1\uparrow},~p_{L\uparrow}$ are the ones given in
(\ref{pb}).
The energy eigenvalue $E$ of the  model is given by
$E=-2\sum ^N_{j=1}\cos k_j$ (up to some additive constants,
which we have dropped).

\vskip.3in
%\acknowledgments
Y.-Z.Z is supported by the QEII Fellowship Grant from
Australian Research Council. 

\appendix

%\newpage
%\vskip.3in

\end{document}